# Antiferromagnetism and phase transitions in non-centrosymmetric UIrSi$_3$

J. Valenta[1], F. Honda[2], M. Vališka[1], P. Opletal[1], J. Kaštil[3], M. Míšek[3], M. Diviš[1], L. Sandratskii[4], J. Prchal[1], and V. Sechovský[1]

[1]*Charles University, Faculty of Mathematics and Physics, Department of Condensed Matter Physics, Ke Karlovu 5, Prague 2, Czech Republic*
[2]*Tohoku University, Institute for Materials Research, Narita-cho 2145-2, Oarai, Ibaraki, Japan*
[3]*Institute of Physics AS CR, Na Slovance 1999/2, Prague 8, Czech Republic*
[4]*Max-Planck-Institute of Microstructure Physics, Weinberg 2, 06120 Halle, Germany*

**Abstract**

Magnetization and specific heat measurements on a UIrSi$_3$ single crystal reveal Ising-like antiferromagnetism below $T_N$ = 41.7 K with easy magnetization direction along the *c*-axis of tetragonal structure. The antiferromagentic ordering is suppressed by magnetic fields > $H_c$ ($\mu_0 H_c$ = 7.3 T at 2 K) applied along the *c*-axis. The first-order metamagnetic transition at $H_c$ exhibits asymmetric hysteresis reflecting a slow reentry of the complex ground-state antiferromagnetic structure with decreasing field. The hysteresis narrows with increasing temperature and vanishes at 28 K. A second-order metamagnetic transition is observed at higher temperatures. The point of change of the order of transition in the established *H-T* magnetic phase diagram is considered as the tricritical point (at $T_{tc}$ = 28 K and $\mu_0 H_{tc}$ = 5.8 T). The modified-Curie-Weiss-law fits of temperature dependence of the *a*- and *c*-axis susceptibility provide opposite signs of Weiss temperatures, $\Theta_p^a$ ~ -51 K and $\Theta_p^c$ ~ +38 K, respectively. This result and the small value of $\mu_0 H_c$ contrasting to the high $T_N$ indicate competing ferromagnetic and antiferromagnetic interactions responsible for the complex antiferromagnetic ground state. The simultaneous electronic-structure calculations focused on the total energy of ferromagentic and various antiferromagnetic states, the U magnetic moment and magnetocrystalline anisotropy provide results consistent with experimental findings and the suggested physical picture of the system.

## Introduction

The growing interest in materials adopting crystal structures which have no center of symmetry was boosted by the discovery of unconventional superconductivity in CePt$_3$Si[1]. The absence of a center of inversion in the crystal structure along with a Rashba-type antisymmetric spin-orbit (s-o) coupling[2, 3] leads to the possibility of a superconducting state with an admixture of spin-triplet and spin-singlet pairs[4]. The Rashba s-o coupling in materials crystallizing in a noncentrosymmetric crystal structure also causes spin-splitting of the Fermi surface into two Fermi surfaces, which has many intriguing implications in various branches of physics including magnetism[5].

The BaNiSn$_3$-type structure (*I4mm*) illustrated in Fig. 1 is one of the ternary variants of the BaAl$_4$ tetragonal structure. It is adopted by several *RTX*$_3$ compounds (*R*: rare earth, *T*: transition metal, *X*: *p*-electron element). The *R* atoms occupy the corners and the body center of the tetragonal structure whereas the *T-X* sublattice is non-centrosymmetric. The lack of an inversion center in the crystal structure brings about a nonuniform lattice potential *V(r)* along the *c*-axis, whereas the nonuniform lattice potential within the *a-b* plane is canceled out[6].
The *RTX*$_3$ compounds with Ce are of high research interest because they exhibit diverse interesting phenomena like superconductivity with a high critical field, pressure induced superconductivity near a quantum critical point, coexistence of antiferromagnetism and superconductivity, vibron states, etc.[7-12]. The magnetic ordering in these materials is usually antiferromagnetic (AF) with complex propagation vectors[13-16] which indicate competing ferromagnetic and antiferromagnetic exchange interactions.

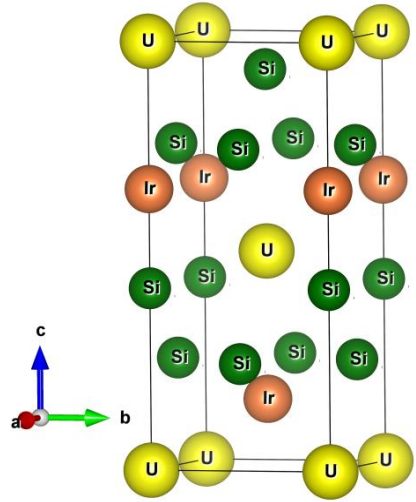

FIG. 1. Crystal structure of UIrSi$_3$.

Contrary to rare-earth compounds where the magnetic moment is usually born in the localized 4*f*-electrons, the 5*f*-electron wave functions in U intermetallics lose, to a considerable extent, their atomic character due to the mutual overlap between neighboring U ions and due to the hybridization of 5*f*-states with valence electron states of ligands (5*f*-ligand hybridization). The large direct overlap of 5*f*-wave functions by rule prevents formation of a rigid atomic 5*f*-electron magnetic moment in materials in which the distance of nearest-neighbor U atoms is smaller than the Hill limit (340-360 pm)[17]. On the other hand, the layout of U-U nearest neighbors carrying 5*f*-electron orbital moments in the crystal lattice usually determines the huge magnetocrystalline anisotropy with the easy magnetization axis perpendicular to the strong U-U bonding planes or chains[18]. The 5$f$-ligand hybridization has similar but more subtle effects on U magnetism. Its role strengthens in compounds with a lower U content in which the U ion surrounding ligands prevent the direct U-U bonds[19, 20]. As concerns the magnetic coupling the direct overlap of 5$f$-wave functions of U neighbors is responsible for the direct exchange interaction between U nearest-neighbor magnetic moments whereas the 5$f$-ligand hybridization mediates the indirect exchange interaction between moments of U ions neighboring the involved ligand.

Only two uranium compounds adopting the tetragonal BaNiSn$_3$-type structure are known, namely, UIrSi$_3$ and UNiGa$_3$. Both have been studied in the form of polycrystals only and reported to order antiferromagnetically below 42 K (UIrSi$_3$)[21] and 39 K (UNiGa$_3$)[22], respectively.

This paper is dedicated to the result of our effort to advance understanding of the physics of one of these two compounds. We have prepared a UIrSi$_3$ single crystal, characterized its composition and crystal structure and measured the magnetization and specific heat in a wide range of temperatures and external magnetic fields.

The results confirm that UIrSi$_3$ becomes antiferromagnetic below the Néel temperature $T_N$ = 41.7 K with strongly anisotropic response to an external magnetic field. In the magnetic field along the $c$-axis it undergoes a metamagnetic transition (MT) at a critical field $H_c$ ($\mu_0 H_c$ = 7.3 T at 2 K) into a field-polarized state with a magnetic moment of ~ 0.66 $\mu_B$/f.u. The observed $H_c$ value is much lower than expected for a simple antiferromagnet consisting of magnetic moments of the order of 1 $\mu_B$ with $T_N$ > 40 K. No MT shows up in the $a$-axis field up to 14 T.

At low temperatures, MT is a first order magnetic phase transition (FOMPT) and shows an asymmetric hysteresis. $H_c$ decreases with increasing temperature while the hysteresis shrinks with increasing temperature and eventually vanishes at 28 K. The character of MT dramatically changes at this temperature from FOMPT to a second order magnetic phase transition (SOMPT) which is observed for 28 K < $T$ < $T_N$) as manifested by the change of character of magnetization and specific-heat anomalies.

To understand the observed phenomena further we have performed first-principles electronic structure calculations focused on magnetism in UIrSi$_3$. The corresponding results of experiments and calculations as concerns the type of anisotropy and the magnitude of anisotropy energy show reasonable agreement whereas the agreement on the size of the U magnetic moment depends on the calculation method. Both the experiment and theory suggest that the magnetically compensated ground-state of the system is not a simple two-sublattice antiferromagnet of up-down- up-down type but has a more complex nature.

**Experimental and Computational details**

The process of preparation of a UIrSi$_3$ single crystal was started by synthesis of a stoichiometric polycrystal from pure elements and casting a rod (⌀ 6.5 mm, length 85 mm). High purity elements: U (99.9%), Ir (99.99%) and Si (99.999%) were used. The obtained ingot was mounted in a four-mirror optical furnace (by Crystal Systems Corporation) optimized for the floating zone melting method. The middle part of the final product was annealed at 700°C for 10 days. Energy dispersive x-ray analysis confirmed the presence of the single UIrSi$_3$ phase in the annealed product. The lattice parameters $a$ = 417.22 pm and $c$ = 996.04 pm of the tetragonal BaNiSn$_3$-type structure determined by the x-ray powder diffraction on a pulverized piece of the single crystal are in reasonable agreement with the literature [21]. The U ions are coordinated solely within uranium basal plane layers. Each U ion has four U nearest neighbors located within the same basal plane and separated by $d_{U-U}$ = 417.22 pm ( = $a$).

The magnetization and specific-heat measurements were carried out with a PPMS apparatus (Quantum Design Inc.) in magnetic fields applied along the *c*-axis up to 14 T. For determination of $T_N$ from the temperature dependence of specific heat, the point of the balance of the entropy released at the phase transition was taken. The field dependence of the specific heat was measured point by point in a stable magnetic field. At each point the measurement was repeated four times.

The magnetic moments, easy magnetization axis, magnetocrystalline anisotropy energy, equilibrium volume and stability of antiferromagnetic structures were calculated using the methods based on density functional theory (DFT). We used the computer codes full potential local orbitals (FPLO)[23], full potential augmented plane waves plus local orbitals (APW+lo)[24] and in-house augmented spherical waves (ASW)[25] to solve single particle Kohn-Sham equations. We treated the *5f* states as itinerant Bloch states in all three methods and since no information about ground state magnetic structure is available the simple ferromagnetic and antiferromagnetic arrangements of moments were applied. The fully relativistic Dirac four-component mode was used in all FPLO calculations. For calculations with FPLO code we used the division 24×24×24 for both the *a*- and *c*-axes corresponding to 1764 and 3756 irreducible k-points in the Brillouin zone, respectively, to ensure the convergence of results. Since the total magnetic moment obtained from relativistic calculations was too small in comparison with the experimental one the orbital polarization correction[26] was applied in the FPLO code. In the APW+lo method we applied spin orbit coupling with the local spin density (LSDA) + Hubbard *U* approach[24] to resolve the problem of the small calculated total magnetic moment which points to additional electron correlations beyond the local and semilocal exchange correlation potentials.

The APW+lo method was used to determine equilibrium lattice parameters. We used more than 800 augmented plane waves (more than 160 APWs per atom) and 2000 k-points in the Brillouin zone to obtained converged results. The calculations of equilibrium volume with the APW+lo method were scalar relativistic to use forces when calculated with local spin density (LSDA)[27] and the general gradient approximations (GGA)[28-30].

The ASW-LSDA method including the spin-orbit coupling (SOC) was applied to calculate the total energy difference of the ferromagnetic and the two types of the antiferromagnetic structures. The ASW method is well suited for calculation of complex magnetic structures in Uranium compounds (see, e.g. in Ref.[31])

The performances of LSDA and GGA have been compared with respect to the equilibrium volume of UIrSi$_3$. The experimental *c*/*a* ratio and the symmetry free structure parameters obtained from minimization of the forces were used in all the calculations. We have calculated the variation of the total energy with the relative volume $V/V_0$ ($V_0$ is the experimental equilibrium volume). The LSDA[27] value of the equilibrium volume is about 3.3 % smaller than the experimental value. This is a typical deviation usually obtained in LSDA calculations. The GGA from Ref.[28], on the other hand, leads to a volume that exceeds the experimental $V_0$ by 1.7 % and the volume obtained with the GGA from Ref.[29] is 1.5 % smaller. The best results are obtained using the GGA from Ref.[30], which underestimates $V_0$ by only 0.8 %. In all forms GGA[28-30] provides a better equilibrium volume than LSDA.

**Results and Discussion**

The temperature dependence of the specific heat $C_p(T)$ of UIrSi$_3$ displayed in Fig. 2 exhibits an anomaly at 41.7 K of a lambda shape, characteristic for a second-order phase transition. The anomaly is progressively shifted to lower temperatures when the crystal is subjected to a gradually increasing magnetic field applied along the *c*-axis (see Fig. 3). This is a typical behavior of antiferromagnets at magnetic ordering transition. In analogy we conclude that UIrSi$_3$ in zero magnetic field undergoes a magnetic phase transition between a paramagnetic and an AF state at Néel temperature $T_N$ = 41.7 K which confirms the only published report on this compound[21]. When the magnetic field increases above 4.1 T the height of the anomaly increases and simultaneously the anomaly becomes sharper. This evolution terminates at 5 T. In fields increasing beyond 5 T the peak becomes gradually lower and broader, disappearing around 7.1 T.

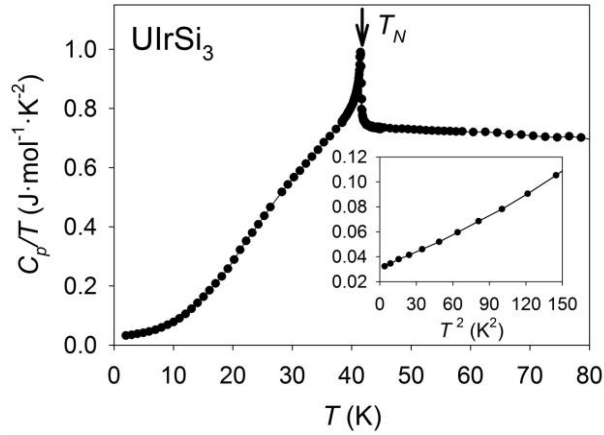

FIG. 2. Temperature dependence of the specific heat of UIrSi$_3$ in the temperature range 2-80 K. The inset shows the low temperature specific heat in the $C_p/T$ vs. $T^2$ plot. The arrow marks $T_N$ = 41.7 K

We have also measured temperature responses to the applied thermal pulse in both, increasing and decreasing temperature regimes. In fields higher than 5.8 T the transition exhibits a temperature hysteresis which indicates emerging latent heat which is accompanying a first-order magnetic phase transition. The hysteresis of the transition vanishes in fields < 5.8 T.

When closely inspecting the specific heat (see Fig. 3) one can observe that in higher fields the peak in the $C_p/T$ vs. *T* plot has no lambda shape seen in fields above 5.6 T anymore.

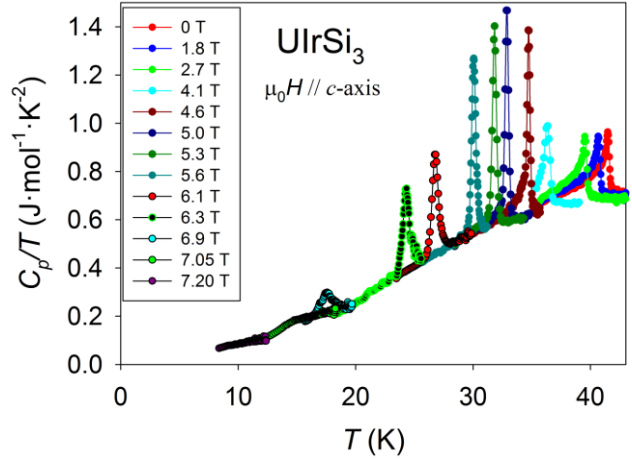

FIG. 3. Temperature dependence of specific heat ($C_p/T$ vs. $T$ plot) of single crystalline UIrSi$_3$ in various magnetic fields applied along the *c*-axis.

In contrast to the pronounced *c*-axis field influence on the specific heat, application of magnetic fields up to 14 T applied along the *a*-axis leaves the entire $C_p(T)$ dependence intact. This indicates uniaxial magnetic anisotropy with the *a*-axis as a hard axis.

The estimated magnetic entropy associated with the AF transition in zero field is very low, namely 0.14Rln2. Such small value is usually attributed to the itinerant character of magnetism. However, the nearest U neighbor ions in UIrSi$_3$ are 417 pm apart which prevents considerable delocalization of 5*f*-electron states due to the overlap of 5*f*-electron wave-functions. The $C_p/T$ vs. $T^2$ plot of low temperature ($T < 4$K) specific-heat data shown in the inset of Fig. 2 is almost linear and points to a value of the Sommerfeld coefficient $\gamma = 31$ mJ·mol$^{-1}$·K$^{-2}$ which is one of the lowest values among U intermetallics.

Another evidence of the AF transition of UIrSi$_3$ at $T_N$ is provided by measurements of the temperature dependence of the magnetic susceptibility $\chi$ (= $M/H$; *M*: magnetization, *H*: magnetic field) displayed in an $M/H$ vs. *T* plot in Fig. 4. One can see that $T_N$ determined from specific heat data falls to a somewhat lower temperature than the maximum of $\chi$ vs. *T* curve. This is because $T_N$ is to coincide with the maximum of the $\partial(\chi T)/\partial T$ derived from the temperature dependence of the genuine thermodynamic variable $\chi T$ [32, 33]. The susceptibility measured in the magnetic field applied along the *c*-axis is much larger than that in the *a*-axis field. Neither shift of the transition temperature nor change of character of the $\chi(T)$ curve are observed for the magnetic field applied along *a*-axis. On the other hand the $\chi(T)$ peak near $T_N$ is gradually shifted to lower temperatures and broadens if the field applied along the *c*-axis is increasing and disappears in a field above 7 T. This behavior correlates with the properties of the specific-heat anomaly related to the AF transition.

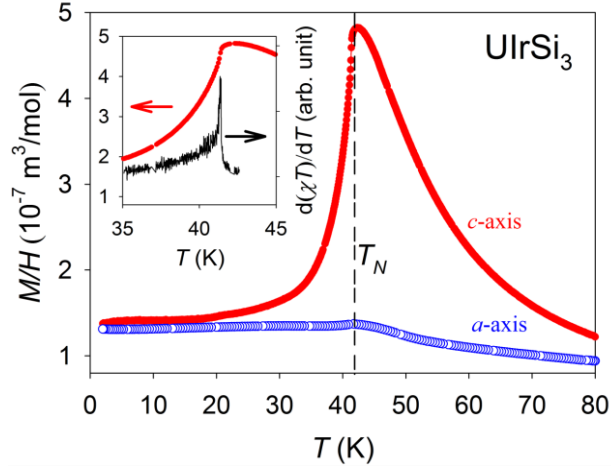

FIG. 4. Temperature dependence of susceptibility (*M/H* vs. *T* plot) of single crystalline UIrSi$_3$ in the temperature range 2-80 K in the magnetic field applied along the *c*-axis ($\mu_0 H$ = 0.1 T) – full red circles, and along the *a*-axis ($\mu_0 H$ = 0.5 T) – empty blue circles. The broken line marks $T_N$ (= 41.7 K) determined from specific heat data whereas the maximum of the *M/H* is at higher temperature (= 42.3 K). Inset: Detail of the *c*-axis *M/H* vs. *T* plot between 35 and 45 K. The full black line represents the function $\partial(\chi T)/\partial T$ vs. *T* dependence. The Néel temperature is supposed to be at the maximum of $\partial(\chi T)/\partial T$ (= 41.4 K).

The zero-field cooled (ZFC) and field cooled (FC) thermomagnetic curves in the *c*-axis field up to 7 T are entirely merging. In fields above 7 T the ZFC and FC curves separate at low temperatures which indicate destabilization of the AF state by the field.

The temperature dependences of the inverse magnetic susceptibility in the paramagnetic region plotted in Fig. 5 demonstrate ubiquity of the strong magnetocrystalline anisotropy which is a common feature of most of magnetic uranium compounds. The magnetocrystalline anisotropy in the paramagnetic range is usually manifested by the difference of Weiss temperatures (paramagnetic Curie temperatures) $\Theta_p$ as parameters of fits of measured susceptibility vs. temperature in the magnetic field applied along the main crystallographic axes by a modified Curie-Weiss (MCW) law:

$$\chi(T) = \frac{C}{T - \theta_P} + \chi_0 , \qquad (1)$$

where *T* is temperature, *C* is the Curie constant from which the value of effective moment $\mu_{\text{eff}}$ can be derived and $\chi_0$ is a temperature independent term representing the susceptibility of conduction electrons.

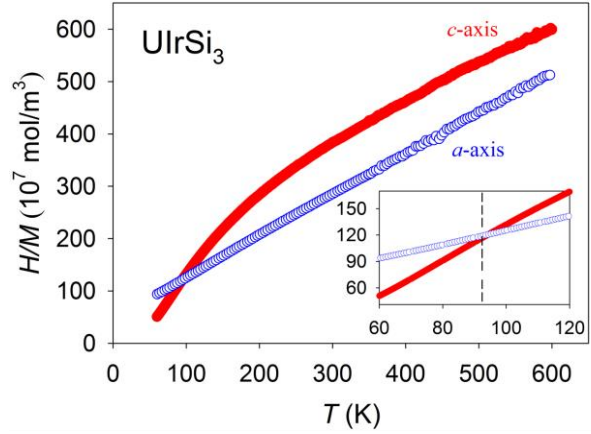

FIG. 5. Temperature dependence of the inverse susceptibility ($H/M$ vs. $T$ plot) of single crystalline UIrSi$_3$ in the temperature range 60-600 K in the magnetic field ($\mu_0 H = 8$ T) applied along the *c*-axis (full red circles) and along the *a*-axis (empty blue circles). The lines representing fits according to formula in the text are hidden by experimental points. The inset shows the 60-120 K detail. The vertical broken line marks the temperature of the crossing of the *a*-axis and *c*-axis $H/M$ vs. $T$ curves.

We have measured the susceptibility in the temperature range from 2 to 600 K but fitted data only above 60 K which is sufficiently higher than $T_N$ to avoid influence of correlations in the proximity of the magnetic ordering transition. Presuming negligible contributions to $\mu_{eff}$ from Ir and Si the fitted effective moment is related to one U ion.

TABLE I. Results of the modified Curie-Weiss law fits of susceptibility data measured on a UIrSi$_3$ single crystal in the magnetic field along the *a*- and *c*-axis.

| $H \parallel$ | fit $T$ range (K) | $\mu_{eff}$ ($\mu_B$/U) | $\Theta_p$ (K) | $\chi_0$ ($10^{-9}$ m$^3$/mol) |
|---|---|---|---|---|
| *a* | 60 – 600 | 2.7 | -51 | 1.3 |
| *c* | 200 – 600 | 2.0 | -24 | 6.5 |
| *c* | 60 – 200 | 1.6 | 38 | 8.3 |

The *a*-axis susceptibility data can be well fitted over the entire temperature range 60 – 600 K. The fitted $\Theta_p$ value of –51 K compares with $T_N$ (41.7 K) as expected in simple antiferromagnets entirely governed by an AF interaction. The fitted $\mu_{eff}$ values are much lower than the expectation values for the U$^{3+}$ and U$^{4+}$ free ion (3.62 $\mu_B$ and 3.58 $\mu_B$, respectively).

On the other hand, the *c*-axis $\chi(T)$ data cannot be fitted by a MCW law over the entire temperature interval. To get some qualitative insight into the complex situation we attempted to fit the data sets in the 200 – 600 K and 60 – 200 K sections separately. The obtained fitting parameters $\Theta_p$, $\mu_{eff}$, and $\chi_0$ are displayed in TABLE I.

Further on we refer to data obtained below 200 K only. The temperature dependencies of the *a*- and *c*-axis susceptibility, respectively, cross at ~ 93 K. The Weiss temperatures $\Theta_p^a$ ~ - 51 K and $\Theta_p^c \sim +38$ K obtained from modified-Curie-Weiss-law fits in the low-temperature region reflect competing ferromagnetic and AF interactions. It is worth mentioning that a qualitatively similar situation (strong concave curvature of the $1/\chi_c$ vs. $T$, crossing of the $1/\chi_a$ vs. $T$ and $1/\chi_c$ vs. $T$ dependences, positive $\Theta_p^c$ vs. negative $\Theta_p^a$ value at low temperatures) is observed in the case of $UIr_2Si_2$[34] crystallizing in the tetragonal $CaBe_2Ge_2$-type structure in which the U ions appear in a noncentrosymmetric surrounding of ligands similar to that in $UIrSi_3$. At this stage of research we have no explanation for the origin of the anisotropy of temperature independent parameter $\chi_0$.

The 2 K magnetization curves measured in the field applied along the $a$- and $c$-axis, respectively, which are displayed in Fig. 6, clearly demonstrate the strong uniaxial anisotropy of $UIrSi_3$ in the AF state with the $c$-axis as the easy magnetization direction. The weak linear $M(H)$ dependence of the magnetization reaching only 0.2 $\mu_B$/f.u. in 14 T is characteristic for the field applied along the $a$-axis, the magnetically hard direction, indicated already by specific-heat measurements. The $c$-axis magnetization follows almost the same $M(H)$ dependence in the field up to 7 T. When the field is increased above 7 T a sharp metamagnetic transition (MT) emerges resulting in a magnetization step up to 0.66 $\mu_B$.

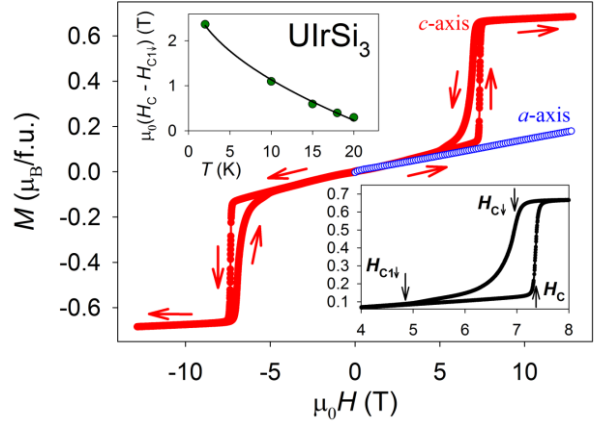

FIG. 6. The hysteresis loop of the magnetization measured at 2 K in the magnetic field applied along the $c$-axis (full red circles) and the $M$ vs. $H$ dependence of the magnetization along the $a$-axis (empty blue circles). The arrows show the polarity of the field sweep.
Bottom right inset: The $\mu_0 H$ = 4-8 T detail in the magnetic field applied along the $c$-axis. Top left inset: Temperature evolution of the hysteresis of the metamagnetic transition. The line represents the fit of experimental points with formula (2) in the text.

The values of spin and orbital magnetic moments of uranium were calculated for a ferromagnetic configuration from Dirac-Kohn-Sham equations and the values $M_s$ = -1.42 and $M_L$ = 1.87 $\mu_B$ providing the value of total moment $M$ = 0.45 $\mu_B$. The total moment is much smaller than experimentally determined moment of 0.66 $\mu_B$ in a field above MT. The orbital polarization correction as implemented in FPLO code[26] has been applied providing a total magnetic moment 1.78 $\mu_B$ which overestimates the experimental value. Therefore the relativistic $LSDA+U$ method with adjustable Hubbard parameter $U$ was applied[24]. The spin $M_s$ = -1.20 $\mu_B$ and orbital $M_L$ = 1.86 $\mu_B$ magnetic moments were calculated with the parameter $U$ = 0.28 eV that gives very good agreement with experimental saturated moment. This finding showed the 5$f$ electrons in $UIrSi_3$ are moderately correlated. The moments calculated by the various methods and the experimentally determined moment are compared in TABLE II.

TABLE II. The spin $M_S$, orbital $M_L$ and total magnetic moments $M$ calculated by different methods. Experimental value M means saturated moment.

| Method | $M_S$ [$\mu_B$] | $M_L$ [$\mu_B$] | $M$ [$\mu_B$] |
|---|---|---|---|
| *Dirac* | -1.42 | 1.87 | 0.45 |
| *Dirac + OPC* | -1.99 | 3.77 | 1.78 |
| *LSDA + U + SOC* | -1.20 | 1.86 | 0.66 |
| Experiment | | | 0.66 |

The total energies of the ferromagnetic and simple antiferromagnetic (up-down-up-down...) structures calculated by the ASW-LSDA method including SOC were compared. The AFM structure was formed by the ferromagnetic U layers in the a-b planes. The subsequent layers had opposite directions of the magnetic moments. The magnetic moments were collinear to the easy *c*-axis. Such an AFM structure appeared to be higher in energy than the FM structure by 16 meV/f.u. We increased the supercell along the c axis and tried an up-up-down-down antiferromagnetic structure. Interestingly, in this case there are two types of inequivalent U ions but the structure remains magnetically compensated. The energy of this structure appeared to be much closer to the energy of the FM one although still higher than the FM energy by about 5 meV/f.u. These results are in line with our expectations of a complex antiferromagnetic ordering in UIrSi$_3$ due to competition of ferromagnetic and antiferromagnetic exchange interactions. The theoretical study of complex magnetically compensated structures in UIrSi$_3$ will be continued.

When we extrapolate the *a*- and *c*-axis magnetization curves beyond the maximum applied magnetic field we find them crossing at ~59 T which serves as a rough estimate of a magnetocrystalline anisotropy field. This value is about an order of magnitude smaller than the anisotropy fields of the majority of U*TX* and U*T*$_2$*X*$_2$ compounds which typically exhibit anisotropy fields of several hundred teslas[18]. On the other hand it compares to the anisotropy fields of UIr$_2$Si$_2$ and UPt$_2$Si$_2$ which both crystallize in the CaBe$_2$Ge$_2$-type structure[34, 35].

To estimate magnetocrystalline anisotropy energy (MAE) the total energy with magnetic moment along *a*- and *c*-axes, respectively, have been calculated. The *c*-axis was found to be the easy axis and the MAE is 4.57 meV which is in fair agreement with the experimental one ~1.1 meV.

The critical field $H_c$ of FOMPT is taken as the midpoint of the magnetization step when sweeping the field up. Hysteresis is an intrinsic characteristic of first order transitions [36]. We indeed have observed a hysteresis although rather unusual. The reverse (field-sweep-down) transition is considerably broader and nonsymmetric due to a tail in low fields. We tentatively attribute this behavior to gradual reentry of a complex ground-state AF spin arrangement with intermediate uncompensated phases. This picture should be verified by a relevant microscopic experiment (neutron scattering). To characterize this transition, we have introduced the characteristic fields $H_{c\downarrow}$ and $H_{c1\downarrow}$ as indicated in the inset of Fig. 6.

The temperature evolution of MT is demonstrated in Fig. 7 where the magnetization isotherms *M(H)* measured at selected temperatures are shown. $H_c$ decreases with increasing temperature whereas the magnetization step at MT and the hysteresis get reduced. The temperature dependence of hysteresis $\mu_0(H_c - H_{c1\downarrow})$ is shown in Fig. 6 and can be well fitted by the formula

$$\mu_0(H_c - H_{c1\downarrow}) = c\left(1 - \sqrt{\frac{T}{T_1}}\right), \qquad (2),$$

which has been taken *ad hoc* from the paper on the temperature dependence of the coercive field in single-domain particle systems applied to the $Cu_{97}Co_3$ and $Cu_{90}Co_{10}$ granular alloys[37]. The fitting parameters $c = 3.3$ and $T_1 = 24$ K.

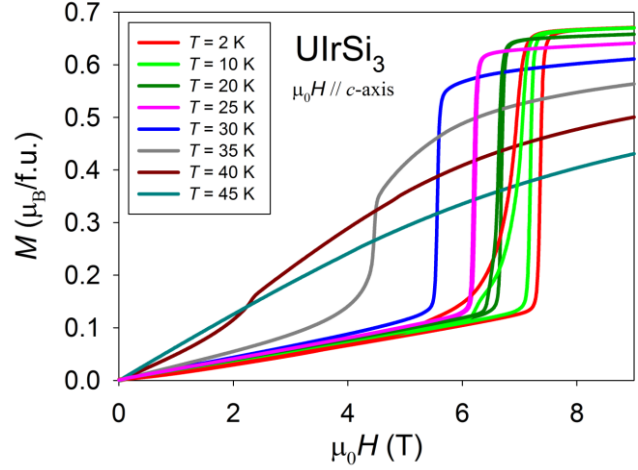

FIG. 7. The magnetization curves measured at various temperatures in the magnetic field applied along the *c*-axis.

Two types of *M(H)* curves can be seen in Figs. 7 – 9:

i) at $T \leq 25$ K they are characterized by a magnetization step $\Delta M$ at $H_c$ and a field hysteresis $\Delta H = (H_c - H_{c1\downarrow})$ and are practically linear for $H < 0.9\, H_c$ and saturated for $H > 1.1\, H_c$ and $H_c$, $\Delta M$ and $\Delta H$ decrease with increasing temperature. These attributes are characteristics of a FOMPT,

ii) at temperatures $T_N \geq T \geq 30$ K the *M(H)* curves show an upturn in fields $\leq H_c$ terminated by a cusp at $H_c$ and followed by gradual saturation in higher fields. $H_c$ decreases with increasing temperature to become 0 T at $T_N$. The upturn and the cusp become simultaneously less pronounced and no hysteresis is observed. In our scenario these features are characteristic of a SOMPT.

The magnetic-field dependences of specific heat are shown in Figs. 8 and 9. The plots presented in Figs. 8, and 9 manifest the dramatic difference between the effects in the specific heat accompanying the FOMPT (Fig. 8) and the SOMPT (Fig. 9), respectively, and corresponding magnetization behavior. The FOMPT is manifested by a step of $C_p/T$ at $H_c$ (and hysteresis which qualitatively resembles magnetization behavior. The positive step of $C_p/T$ at $H_c$ is understood as a result of an increased density of conduction electron states due to reconstruction of the Fermi surface at MT.

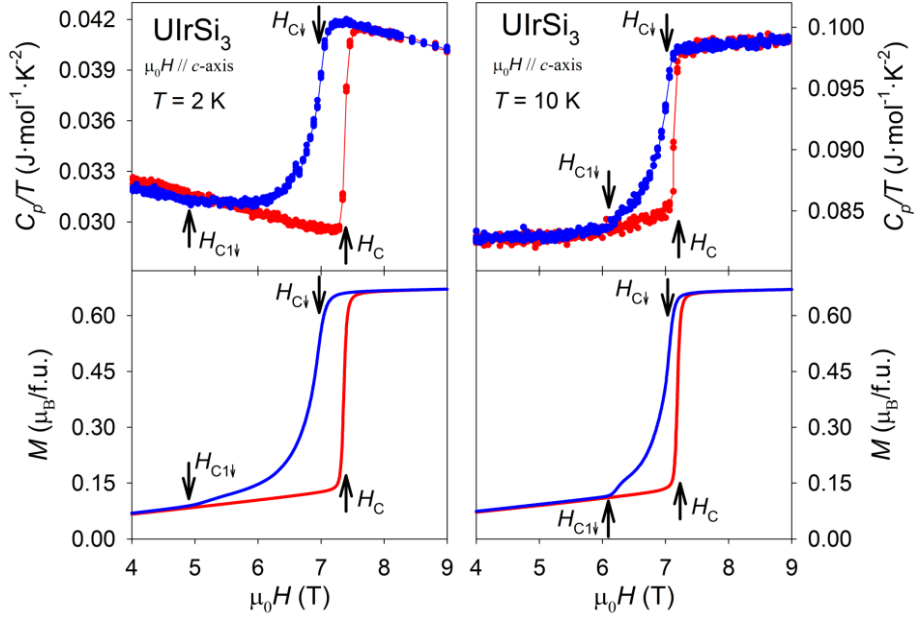

FIG. 8. Magnetic field dependence of specific heat ($C_p/T$ vs. $\mu_0 H$ plot) and magnetization of single crystalline UIrSi$_3$ at 2 K (left panel) and 10 K (right panel) in the magnetic field applied along the $c$-axis (field sweep up - red, field sweep down - blue).

On the other hand, the SOMPT is manifested by a $\lambda$-shape anomaly at $H_c$. The enhanced $C_p/T$ values in lower fields reflect spin-flip fluctuations from the AF state. The enhancement is more pronounced with temperature approaching $T_N$.

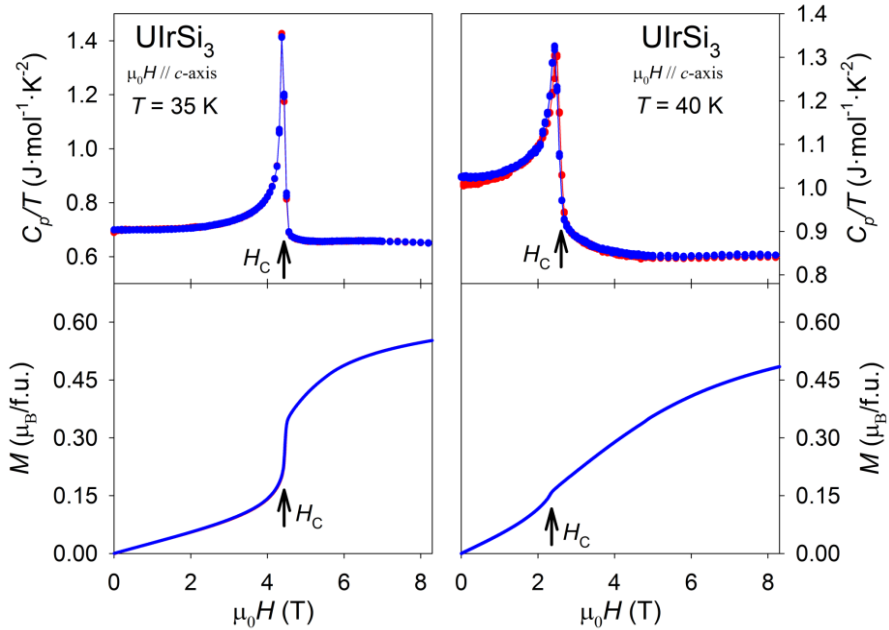

FIG. 9. Magnetic field dependence of specific heat ($C_p/T$ vs. $\mu_0 H$ plot) and magnetization of single crystalline UIrSi$_3$ at 35 K (left panel) and 40 K (right panel) in the magnetic field applied along the $c$-axis (field sweep up - red, field sweep down - blue).

The magnetization and specific-heat data allowed us to establish the $H$-$T$ magnetic phase diagram shown Fig. 10. The critical field of the MT decreases with increasing temperature towards zero at $T_N$. UIrSi$_3$ undergoes a FOMPT at temperatures $T < 28$ K contrary to a SOMPT observed for $T > 28$ K. The strikingly different magnetization response to fields below $H_c$ in the low and elevated temperature areas of the magnetic phase diagram evokes a question whether also the corresponding AF phases are different. We tentatively consider the point separating the FOMPT and SOMPT regimes in the magnetic phase diagram as a tricritical point with coordinates $T_{tc} = 28$ K, $\mu_0 H_{tc} = 5.8$ T. Recently two papers reported similar to us a tricritical point in uranium intermetallic antiferromagnets without closer specification of the three involved phases[38, 39]. We are fully aware of the weakness of the suggested scenario until the difference of the two antiferromagnetic phases is proven by relevant experimental methods, e.g. neutron scattering, μSR, etc.

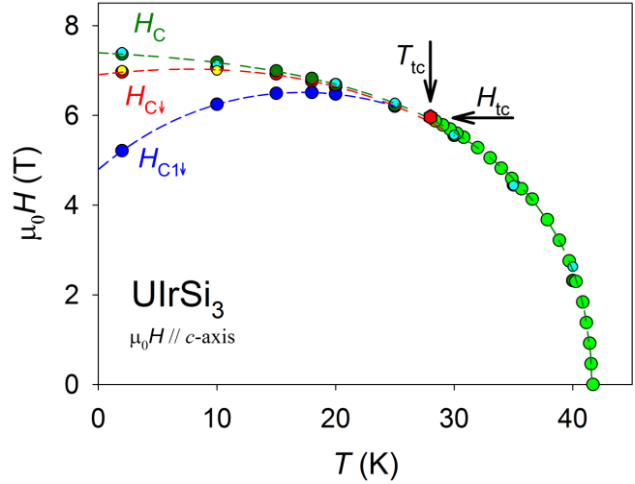

FIG. 10. Magnetic phase diagram of UIrSi$_3$ in the magnetic field applied along the *c*-axis. The labels $H_c$, $H_{c\downarrow}$ and $H_{c1\downarrow}$ are defined in the inset of Fig. 6. Colors of data points are representing data from measurements of: $M(H)$ … $H_c$ - dark green, $H_{c\downarrow}$ - red, $H_{c1\downarrow}$ - blue; $C_p(H)$ … $H_c$ - light blue, $H_{c\downarrow}$ - yellow; $C_p(T)$... $H_c$ - light green followed by a red hexagon indicating the tricritical point ($T_{tc} = 28$ K, $H_{tc} = 5.8$ T).

**Conclusions**

We have grown a single crystal of the noncentrosymmetric tetragonal UIrSi$_3$ compound. The UIrSi$_3$ stoichiometry and the BaNiSn$_3$-type structure have been confirmed by EDX and x-ray diffraction analysis, respectively. The crystal was subjected to detailed measurements of magnetization and specific heat with respect to temperature and external magnetic field. To understand the experimental results, more-detailed first-principles electronic structure calculations for this compound have been performed.

The results point to Ising-like antiferromagnetism below $T_N = 41.7$ K exhibiting strong uniaxial anisotropy with the easy magnetization direction along the *c*-axis. The competition of antiferro- and ferromagnetic exchange interactions plays an important role in UIrSi$_3$ magnetism as manifested by the a) different sign of paramagnetic Curie temperatures of the *a*- and *c*-axis susceptibility ($\Theta_p^a \sim -51$ K and $\Theta_p^c \sim +38$ K), b) low critical field of MT ($\mu_0 H_c = 7.3$ T at 2 K) contrasting with high $T_N$ (= 41.7 K), c) asymmetric hysteresis of MT, d) existence of regions characterized by the first and second order phase transition, respectively, separated by a tricritical point (at $T_{tc}= 28$ K, $\mu_0 H_{tc} = 5.8$ T), e) a higher calculated total energy of a simple AF ground state than a complex magnetically compensated state leading to the prediction of a complex AF ground-state magnetic structure. The existence of possible different antiferromagnetic phases remains to be proven by relevant microscopic methods.

**Acknowledgments**

This research is supported by Grant Agency of Charles University (GAUK) by project no. 188115, the Czech Science Foundation, grant No. P204/15-03777S and the Japan Society for the Promotion of Science (JSPS) KAKENHI with the grant nos. JP15K05156 and JP15KK0149. Experiments were performed in the Materials Growth and Measurement Laboratory MGML (http://mgml.eu). The authors are indebted to Dr. Ross Colman for critical reading and correcting of the manuscript.